\documentclass[twocolumn,pre]{revtex4}

\usepackage{dcolumn}
\usepackage{amsmath}

\usepackage{graphicx}
\usepackage{rotating}
\usepackage{multirow}

\begin{document}

\renewcommand{\d}{{\rm d}}
\renewcommand{\i}{{\rm i}}
\renewcommand{\O}{{\rm O}}
\newcommand{\e}{{\rm e}}
\newcommand{\defn}{\textit}
\newcommand{\half}{\mbox{$\frac12$}}
\newcommand{\set}[1]{\lbrace#1\rbrace}
\newcommand{\av}[1]{\langle#1\rangle}
\newcommand{\eref}[1]{(\ref{#1})}
\newcommand{\etal}{{\it{}et~al.}}
\newcommand{\ve}{\mathbf{e}}
\newcommand{\vx}{\mathbf{x}}
\newcommand{\vs}{\mathbf{s}}
\newcommand{\vH}{\mathbf{H}}
\newcommand{\vone}{\mathbf{1}}
\newcommand{\vm}{\mathbf{m}}
\newcommand{\vI}{\mathbf{I}}
\newcommand{\vA}{\mathbf{A}}
\newcommand{\Tr}{\mathop{\rm Tr}}
\newcommand{\cov}{\mathop{\rm cov}}
\newcommand{\var}{\mathop{\rm var}}
\newcommand{\Li}{\mathop{\rm Li}}
\newcommand{\norm}[1]{\|\,#1\,\|}
\newcommand{\phm}{\phantom{-}}

\newlength{\figurewidth}
\ifdim\columnwidth<10.5cm
  \setlength{\figurewidth}{0.95\columnwidth}
\else
  \setlength{\figurewidth}{10cm}
\fi
\setlength{\parskip}{0pt}
\setlength{\tabcolsep}{6pt}

\title{Mixing patterns in networks}
\author{M. E. J. Newman}
\affiliation{Department of Physics, University of Michigan, Ann Arbor,
MI 48109--1120}
\affiliation{Santa Fe Institute, 1399 Hyde Park Road, Santa Fe, NM 87501}
\begin{abstract}
We study assortative mixing in networks, the tendency for vertices in
networks to be connected to other vertices that are like (or unlike) them
in some way.  We consider mixing according to discrete characteristics such
as language or race in social networks and scalar characteristics such as
age.  As a special example of the latter we consider mixing according to
vertex degree, i.e.,~according to the number of connections vertices have
to other vertices: do gregarious people tend to associate with other
gregarious people?  We propose a number of measures of assortative mixing
appropriate to the various mixing types, and apply them to a variety of
real-world networks, showing that assortative mixing is a pervasive
phenomenon found in many networks.  We also propose several models of
assortatively mixed networks, both analytic ones based on generating
function methods, and numerical ones based on Monte Carlo graph generation
techniques.  We use these models to probe the properties of networks as
their level of assortativity is varied.  In the particular case of mixing
by degree, we find strong variation with assortativity in the connectivity
of the network and in the resilience of the network to the removal of
vertices.
\end{abstract}
\pacs{89.75.Hc, 87.23.Ge, 64.60.Ak, 05.90.+m}
\maketitle

\section{Introduction}
\label{intro}
The techniques of statistical physics were developed to study the
properties of systems of many interacting particles, atoms, or molecules,
but their applicability is wider than this, and recent work has fruitfully
applied these techniques to economies, ecosystems, social interactions, the
Internet, and many other systems of current interest.  The component parts
of these systems, the analogs of atoms and molecules, such things as
traders in a market, or computers on the Internet, are not usually
connected together on a regular lattice as the atoms of a crystal are.  Nor
indeed do their patterns of connection normally fit any simple
low-dimensional structure.  Instead they fall on some more generalized
``network,'' which may be more or less random depending on the nature of
the system.  The broadening of the scope of statistical physics to cover
these systems has therefore led us to the consideration of the structure
and function of networks, as one of the fundamental steps to understanding
real-world phenomena of many kinds.  Useful reviews of work in this area
can be found in Refs.~\onlinecite{Strogatz01,AB02,DM02}.

Recent studies of network structure have concentrated on a small number of
properties that appear to be common to many networks and can be expected to
affect the functioning of networked systems in a fundamental way.  Among
these, perhaps the best studied are the ``small-world
effect''~\cite{TM69,WS98}, network transitivity or
``clustering''~\cite{WS98}, and degree distributions~\cite{BA99b,ASBS00}.
Many other properties however have been examined and may be equally
important, at least in some systems.  Examples include resilience to the
deletion of network nodes~\cite{AJB00,Broder00,CEBH00,CNSW00,Holme02a},
navigability or searchability of
networks~\cite{Kleinberg00proc,ALPH01,WDN02}, community
structure~\cite{GN02,HHJ02,Guimera03}, and spectral
properties~\cite{Monasson99,GKK01,FDBV01}.  In this paper we study another
important network feature, the correlations between properties of adjacent
network nodes known in the ecology and epidemiology literature as
``assortative mixing.''

The very simplest representation of a network is a collection of points,
usually called vertices or nodes, joined together in pairs by lines,
usually called edges or links.  More sophisticated network models may
introduce other properties of the vertices or the edges.  Edges for example
may be directed---they point in one particular direction---or may have
weights, lengths, or strengths.  Vertices can also have weights or other
numerical quantities associated with them, or may be drawn from some
discrete set of vertex types.  In the study of social networks, the
patterns of connections between people in a society, it has long been known
that edges do not connect vertices regardless of their property or type.
Patterns of friendship between individuals for example are strongly
affected by the language, race, and age of the individuals in question,
among other things.  If people prefer to associate with others who are like
them, we say that the network shows assortative mixing or assortative
matching.  If they prefer to associate with those who are different it
shows disassortative mixing.  Friendship is usually found to be assortative
by most characteristics.

Assortative mixing can have a profound effect on the structural properties
of a network.  For example, assortative mixing of a network by a discrete
characteristic will tend to break the network up into separate communities.
If people prefer to be friends with others who speak their own language,
for example, then one might expect countries with more than one language to
separate into communities by language.  Assortative mixing by age could
cause stratification of societies along age lines.  And while the main
focus of this paper is on social networks, it is reasonable to suppose that
similar mixing effects are seen in non-social networks also.  We will give
some examples of this in Section~\ref{bydegree}.

In this paper we study assortative mixing of various types using empirical
network data, analytic models, and numerical simulation.  We demonstrate
that assortative (or disassortative) mixing is indeed present in many
networks, show how it can be measured, and examine its effect on network
structure and behavior.  The outline of the paper is as follows.  In
Section~\ref{discrete} we study the effects of assortative mixing by
discrete characteristics such as language or race.  In Section~\ref{scalar}
we study mixing by scalar properties such as age and particularly vertex
degree; since degree is itself a property of the network topology, the
latter type of mixing leads to some novel network structures not seen with
other types.  In Section~\ref{concs} we give our conclusions.  A
preliminary report of some of the results in this paper has appeared
previously as Ref.~\onlinecite{Newman02f}.

\section{Discrete characteristics}
\label{discrete}
In this section we consider assortative mixing according to discrete or
enumerative vertex characteristics.  Such mixing can be characterized by a
quantity $e_{ij}$, which we define to be the fraction of edges in a network
that connect a vertex of type~$i$ to one of type~$j$.  On an undirected
network this quantity is symmetric in its indices $e_{ij}=e_{ji}$, while on
directed networks or bipartite networks it may be asymmetric.  It satisfies
the sum rules
\begin{equation}
\sum_{ij} e_{ij} = 1,\qquad \sum_j e_{ij} = a_i,\qquad \sum_i e_{ij} = b_j,
\label{sumrules}
\end{equation}
where $a_i$ and $b_i$ are the fraction of each type of end of an edge that
is attached to vertices of type~$i$.  On undirected graphs, where the ends
of edges are all of the same type, $a_i=b_i$~\footnote{We find it
convenient even on a undirected graph to consider the ends of the edges to
be distinguishable---each edge has a unique A-end and B-end, which are
marked in some way.  We can think of one of the ends as having a dot or
other identifying feature on it.  This makes the counting of edges simpler:
the matrix element $e_{ij}$ is defined as the probability that a randomly
chosen edge is connected to a vertex of type~$i$ at its A-end and type~$j$
at its B-end.  Thus every edge, whether it joins unlike vertices or like
ones, appears only once in the matrix---no edge appears both above and
below the diagonal.  It is possible to construct a theory in which the ends
of undirected edges are indistinguishable, but in this case each edge that
joins unlike vertices appears twice in the matrix, both above and below the
diagonal, and edges joining like vertices appear only once.  This
necessitates the introduction of an extra factor of~2 into the off-diagonal
terms.  This approach is adopted for example in Ref.~\onlinecite{VW02}.}.

For example, Table~\ref{sanfran} shows data on the values of $e_{ij}$ for
mixing by race among sexual partners in a 1992 study carried out in the
city of San Francisco, California~\cite{CCKF92}.  This part of the study
focused on heterosexuals, so this is a bipartite network, the two vertex
types representing men and women, with edges running only between vertices
of unlike types.  This means that in this case the ends of an edge are
different and the matrix $e_{ij}$ is asymmetric.  As the table shows,
mixing is highly assortative in this network, with individuals strongly
preferring partners from the same group as themselves.

\begin{table}[t]
\begin{tabular}{l|r|cccc|c}
\multicolumn{2}{c|}{} & \multicolumn{4}{c|}{women}               \\
\cline{3-6}
\multicolumn{2}{c|}{} & black & hispanic & white & other & $a_i$ \\
\hline
\begin{rotate}{90}
\hbox{\hspace{-21pt}men}
\end{rotate}
& black    & 0.258 & 0.016 & 0.035 & 0.013 & 0.323 \\
& hispanic & 0.012 & 0.157 & 0.058 & 0.019 & 0.247 \\
& white    & 0.013 & 0.023 & 0.306 & 0.035 & 0.377 \\
& other    & 0.005 & 0.007 & 0.024 & 0.016 & 0.053 \\
\hline
\multicolumn{2}{r|}{$b_i$} 
           & 0.289 & 0.204 & 0.423 & 0.084 \\
\end{tabular}
\caption{The mixing matrix $e_{ij}$ and the values of $a_i$ and $b_i$ for
sexual partnerships in the study of Catania~\etal~\cite{CCKF92}.  After Morris~\cite{Morris95}.}
\label{sanfran}
\end{table}

\subsection{Measuring discrete assortative mixing}
\label{discmeas}
To quantify the level of assortative mixing in a network we define an
\defn{assortativity coefficient} thus:
\begin{equation}
r = {\sum_i e_{ii} - \sum_i a_i b_i\over 1 - \sum_i a_i b_i}
  = {\Tr\ve - \norm{\ve^2}\over 1 - \norm{\ve^2}},
\label{defsr1}
\end{equation}
where $\ve$ is the matrix whose elements are $e_{ij}$ and $\norm{\vx}$
means the sum of all elements of the matrix~$\vx$.  This formula gives
$r=0$ when there is no assortative mixing, since $e_{ij}=a_ib_j$ in that
case, and $r=1$ when there is perfect assortative mixing and $\sum_i
e_{ii}=1$.  If the network is perfectly disassortative, i.e.,~every edge
connects two vertices of different types, then $r$ is negative and has the
value
\begin{equation}
r_{\rm min} = - {\sum_i a_i b_i\over 1 - \sum_i a_i b_i},
\label{defsrmin}
\end{equation}
which lies in general in the range~$-1\le r<0$.  One might ask what this
value signifies.  Why do we not simply have $r=-1$ for a perfectly
disassortative network?  The answer is that a perfectly disassortative
network is normally closer to a randomly mixed network than is a perfectly
assortative network.  When there are several different vertex types
(e.g.,~four in the case shown in Table~\ref{sanfran}) then random mixing will
most often pair unlike vertices, so that the network appears to be mostly
disassortative.  Therefore it is appropriate that the value $r=0$ for the
random network should be closer to that for the perfectly disassortative
network than for the perfectly assortative one.

A quantity with properties similar to those of Eq.~\eref{defsr1} has been
proposed previously by Gupta~\etal~\cite{GAM89}.  However the definition of
Gupta~\etal\ gives misleading results in certain situations, such as, for
example, when one type of vertex is much less numerous than other types, as
is the case in Table~\ref{sanfran}.  In this paper therefore we use
Eq.~\eref{defsr1}, which doesn't suffer from this problem.  The difference
between the two measures is discussed in more detail in
Appendix~\ref{appa}.

Using the values from Table~\ref{sanfran} in Eq.~\eref{defsr1}, we find that
$r=0.621$ for the network of sexual partnerships, implying, as we observed
already, that this network is strongly assortative by race---individuals
draw their partners from their own group far more often than one would
expect on the basis of pure chance.

As another example of the application of Eq.~\eref{defsr1}, consider the
network studied by Girvan and Newman~\cite{GN02} representing the timetable
of American college football games, in which vertices represent
universities and colleges, and edges represent regular season games between
teams during the year in question.  Colleges are grouped into
``conferences,'' which are defined primarily by geography, and teams
normally play more often against other teams in their own conference than
they do against teams from other conferences.  In other words, there should
be assortative mixing of colleges by conference in the schedule network.
For the 2000 season schedule studied in Ref.~\onlinecite{GN02}, we find a
value for the assortativity coefficient of $r=0.586$, again indicating
strong assortative mixing, i.e.,~colleges do indeed play games with their
conference partners to a substantially greater degree than one would expect
in a randomly mixed network.

It is also useful to know the expected statistical error on the value
of~$r$, so that we can evaluate the significance of our results.  One way
to calculate this error is to use the jackknife method~\cite{Efron79}.
Regarding each of the $M$ edges in a network as an independent measurement
of the contributions to the elements of the matrix~$\ve$, we can show that
the expected standard deviation~$\sigma_r$ on the value of~$r$ satisfies
\begin{equation}
\sigma_r^2 = \sum_{i=1}^M (r_i-r)^2,
\label{jackknife}
\end{equation}
where $r_i$ is the value of $r$ for the network in which the $i$th edge is
removed.  For example, in the case of the matrix of Table~\ref{sanfran}
this gives $\sigma_r=0.014$, which, when compared with the value $r=0.621$
shows that our finding of assortative mixing is strongly statistically
significant---a $40\sigma$ result.

Although it has a rather different physical interpretation, the
coefficient~$r$ is mathematically similar to the intraclass correlation
coefficients used in statistics to compare measurements across different
groups in a population~\cite{Fleiss81}.  Standard results for errors on
intraclass correlations can be adapted to the present case to show that
another estimate of the error on~$r$ is~\cite{FCE69}
\begin{equation}
\sigma_r^2 = {1\over M}\,
             {\sum_i a_i b_i + \bigl[\sum_i a_i b_i\bigr]^2
              - \sum_i a_i^2 b_i - \sum_i a_i b_i^2\over
              1 - \sum_i a_i b_i},
\label{fleiss}
\end{equation}
which gives $\sigma_r=0.012$ for the data of Table~\ref{sanfran},
comparable with the jackknife method.  Either method for
estimating~$\sigma_r$ will be adequate for most purposes---the choice
between them is a matter of convenience.

\subsection{Models of discrete assortative networks}
\label{modelsdisc}
The generalized random graph models of networks studied in the past by
various authors~\cite{BC78,Luczak92,MR95,MR98,ACL00,NSW01,NWS02}
can be extended to the case of assortative mixing on discrete
characteristics.  Suppose we are told the degree distribution~$p_k^{(i)}$
for vertices of type $i=1\ldots n$ in a network and the value of the mixing
matrix $e_{ij}$.  Implicitly, we are also told the values of the quantities
$a_i$ and~$b_i$, since we can extract them from $e_{ij}$ using
Eq.~\eref{sumrules}.  We consider the ensemble of all graphs with these
values of $p_k^{(i)}$ and~$e_{ij}$, which gives us a random graph model
similar in spirit to that of
Refs.~\onlinecite{BC78,Luczak92,MR95,MR98,ACL00,NSW01,NWS02} for
the case of specified degree distribution only.  Many properties of this
ensemble can be calculated exactly in the limit of large system size, as we
now demonstrate.

Suppose that a vertex of type~$i$ has degree~$k$.  The $k$ vertices at the
other ends of the edges attached to this vertex are divided among the $n$
possible vertex types according to some partition $\set{r_1,r_2,\ldots
r_n}$, where $\sum_j r_j=k$.  The probability that the partition
$\set{r_j}$ takes a particular value is given by the multinomial
distribution
\begin{equation}
P^{(i)}(k,\set{r_j}) = k! \prod_j {1\over r_j!}
  \biggl[{e_{ij}\over\sum_j e_{ij}}\biggr]^{r_j}.
\end{equation}
Now, generalizing Ref.~\onlinecite{NSW01}, we define a generating function
for the distributions of the numbers of edges connecting to each type of
vertex:
\begin{eqnarray}
G_0^{(i)}(x_1,x_2,\ldots x_n) &=& \nonumber\\
 & & \hspace{-11em} \sum_{k=0}^\infty p_k^{(i)} \sum_{\set{r_j}}
\delta\bigl(k,\sum_j r_j\bigr) P^{(i)}\bigl(k,\set{r_j}\bigr)
x_1^{r_1} x_2^{r_2}\ldots x_n^{r_n}.
\end{eqnarray}
Performing the sum over $\set{r_j}$, this gives
\begin{eqnarray}
G_0^{(i)}(x_1,x_2,\ldots x_n) &=& \sum_k p_k^{(i)} \Biggl[ {\sum_j e_{ij}
  x_j\over\sum_j e_{ij}} \Biggr]^k\nonumber\\ &=& G_0^{(i)}\Biggl( {\sum_j
  e_{ij} x_j\over\sum_j e_{ij}} \Biggr),
\label{defsg0xx}
\end{eqnarray}
where
\begin{equation}
G_0^{(i)}(x) = \sum_k p_k^{(i)} x^k
\end{equation}
is the fundamental generating function for the degree
distribution~$p_k^{(i)}$, as defined in Ref.~\onlinecite{NSW01}.
Similarly, for the edges connected to a vertex of type~$i$ reached by
following a randomly chosen edge on the graph we have
\begin{equation}
G_1^{(i)}(x_1,x_2,\ldots x_n) = G_1^{(i)}\Biggl( {\sum_j e_{ij}
  x_j\over\sum_j e_{ij}} \Biggr),
\label{defsg0x}
\end{equation}
with
\begin{equation}
G_1^{(i)}(x) = {\sum_k k p_k^{(i)} x^{k-1}\over\sum_k k p_k^{(i)}}
             = {1\over z_i} {\d G_0^{(i)}\over\d x}\biggr|_{x=1},
\end{equation}
where $z_i\equiv {G_0^{(i)}}'(1)$ is the mean degree for type~$i$ vertices.

Now we consider the total number of vertices reachable by following an edge
that arrives at a vertex of type~$i$.  This number has a distribution that
is generated by a generating function $H_1^{(i)}$ satisfying a
self-consistency condition of the form
\begin{equation}
H_1^{(i)}(x) = x G_1^{(i)}(H_1^{(1)}(x),\ldots H_1^{(n)}(x)).
\label{defsh1}
\end{equation}
And similarly the distribution of the number of vertices reachable from a
randomly chosen vertex of type $i$ is generated by
\begin{equation}
H_0^{(i)}(x) = x G_0^{(i)}(H_1^{(1)}(x),\ldots H_1^{(n)}(x)).
\end{equation}
By solving these two equations simultaneously, we can derive the complete
component size distribution in our model.  Here however, we will just
calculate some of the more important average statistics of our networks
from the generating functions.  For example, the mean number $s_i$ of
vertices reachable from a vertex of type~$i$ is
\begin{equation}
s_i = {\d H_0^{(i)}\over\!\d x}\biggr|_{x=1}
    = 1 + {G_0^{(i)}}'(1) {\sum_j e_{ij} {H_1^{(j)}}'(1)\over\sum_j e_{ij}}.
\end{equation}
We can write this in matrix form thus:
\begin{equation}
\vs = \vone + \vm_0\cdot\vH_1'(1),
\end{equation}
where $\vm_0$ is the matrix with elements $z_i {e_{ij}/a_i}$.
Differentiating Eq.~\eref{defsh1}, we then find that
\begin{equation}
\vs = \vone + \vm_0\cdot[\vI-\vm_1]^{-1}\cdot\vone,
\label{vsi}
\end{equation}
where $\vone$ is the vector whose elements are all~1, and the matrix
$\vm_1$ has elements
\begin{equation}
[\vm_1]_{ij} = {z_2^{(i)}\over z_1^{(i)}} {e_{ij}\over a_i},
\end{equation}
with $z_1^{(i)}\equiv z_i$ being the mean degree of type~$i$ vertices and
$z_2^{(i)}$ being the mean number of neighbors at distance two from a
type~$i$ vertex.

When the density of edges on the graph is small, $\vs$ in Eq.~\eref{vsi} is
finite, but it grows with increasing density and then diverges when the
determinant of the matrix $\vI-\vm_1$ reaches its first zero.  This point
marks the phase transition at which a giant component first appears in the
network, similar to the phase transition seen in uncorrelated random
graphs.  The condition $\det(\vI-\vm_1)=0$ for the position of the phase
transition is the equivalent of the condition of Molloy and
Reed~\cite{MR95} for the position of the phase transition in an
uncorrelated random graph of arbitrary degree distribution.

The size of the giant component can also be calculated in a straightforward
manner.  We define $u_i$ to be the probability that a vertex of type~$i$,
reached by following a randomly chosen edge in the graph, does not belong
to the giant component.  This probability is precisely equal to the
probability that none of the neighbors of that vertex are themselves
members of the giant component, and hence $u_i$ satisfies the
self-consistency condition
\begin{equation}
u_i = G_1^{(i)}(u_1,\ldots u_n).
\label{discui}
\end{equation}
The probability that a randomly chosen vertex of type~$i$ is not a member
of the giant component is then $G_0^{(i)}(u_1,\ldots u_n)$, and the overall
fraction~$S$ of vertices that are in the giant component is given by
\begin{equation}
S = 1 - \sum_i {a_i\over z_i} G_0^{(i)}(u_1,\ldots u_n),
\label{discgc}
\end{equation}
where we have made use of the fact that the fraction of vertices of
type~$i$ in the network is equal to $a_i/z_i$.

The simultaneous solution of Eqs.~\eref{discui} and~\eref{discgc} gives us
the value of $S$ for our network.  In many cases we find that these
equations are not solvable in closed form, but they can easily be solved
numerically by iteration of Eq.~\eref{discui} from suitable starting values
for the $u_i$, and then substituting the result into~\eref{discgc}.

\subsection{Simulating discrete assortative networks}
\label{simdisc}
We would also like to be able to generate random networks with a given
level of assortative mixing, in order to check our analytical results and
also for use as substrates for other models, such as, for example,
epidemiological models.  A simple algorithm for achieving this is the
following.
\begin{enumerate}
\item We first choose a size for our graph in terms of the number $M$ of
  edges and we draw $M$ edges from the distribution $e_{ij}$.  That is, we
  generate $M$ edges, each one identified by the types of the vertices that
  it connects, in some manner such that the fraction of edges connecting
  vertices of types~$i$ and~$j$ tends to $e_{ij}$ as $M$ becomes large.  In
  practice, a simple transformation method works well~\cite{NB99}.
\item We count the number of ends of edges of each type~$i$, to give the
  sums~$m_i$ of the degrees of vertices in each class.  We calculate the
  expected number~$n_i$ of vertices of each type from $n_i=m_i/z_i$
  (rounded to the nearest integer), where $z_i$ is the desired mean degree
  of vertices of type~$i$.
\item We draw $n_i$ vertices from the desired degree
  distribution~$p_k^{(i)}$ for type~$i$.  In general the degrees of these
  vertices will not sum exactly to~$m_i$ as we want them to.  So we choose
  one vertex at random, discard it, and draw another from the distribution
  $p_k^{(i)}$ until the sum does equal~$m_i$.
\item We pair up the $m_i$ ends of edges of type~$i$ at random with the
  vertices we have generated, so that each vertex has the number of
  attached edges corresponding to its chosen degree.
\item We repeat from step~3 for each vertex type.
\end{enumerate}
This method correctly generates assortatively mixed graphs with the given
$e_{ij}$ in the limit of large graph size.  In Section~\ref{simresults} of
this paper we give some examples of simulations of assortatively mixed
networks.

\begin{figure}
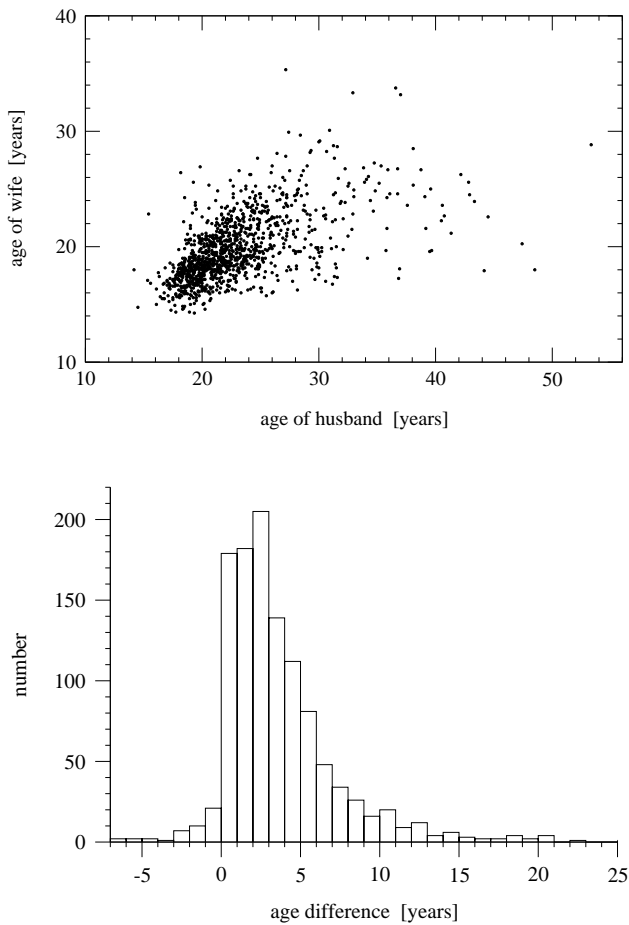

\resizebox{\figurewidth}{!}{\includegraphics{scatter.eps}}
\null\vspace{5ex}
\resizebox{\figurewidth}{!}{\includegraphics{agehist.eps}}
\caption{Top: scatter plot of the ages of 1141 married couples at time of
marriage, from the 1995 US National Survey of Family Growth~\cite{NSFG97}.
Bottom: a histogram of the age differences (male minus female) for the same
data.}
\label{age}
\end{figure}

\section{Assortative mixing by scalar properties}
\label{scalar}
A similar, but distinct, form of assortative mixing is mixing that depends
on one or more scalar properties of network vertices.  A classic example of
mixing of this type seen in many social networks is assortative mixing by
age.  In Fig.~\ref{age} (top panel) we show a scatter plot of the ages of
marriage partners in the 1995 US National Survey of Family
Growth~\cite{NSFG97}.  As is clear from the figure, there is a strong
positive correlation between the ages, with most of the density in the
distribution lying along a rough diagonal in the plot; people, it appears,
prefer to marry others of about the same age, although there is some bias
towards husbands being older than their wives.  In the bottom panel of the
same figure we show a histogram of the age differences in the study, which
emphasizes the same conclusion~\footnote{Perhaps it is stretching a point a
little to regard links between marriage partners as forming a network,
since presumably most people have only one marriage at a time.  However, if
we view the ages of marriage partners as a guide to the ages of sexual
partners in general, then the resulting assortative mixing also describes
networks of such more general partnerships, which are certainly very
real~\cite{GAM89,Morris95,Liljeros01,BMS02}.}.

By analogy with the developments of Section~\ref{discrete}, we can define a
quantity $e_{xy}$, which is the fraction of all edges in the network that
join together vertices with values $x$ and $y$ for the age or other scalar
variable of interest.  The values $x$ and $y$ might be either discrete in
nature (e.g.,~integers, such as age to the nearest year) or continuous
(exact age), making $e_{xy}$ either a matrix or a function of two
continuous variables.  Here, for simplicity, we concentrate on the discrete
case, but generalization to the continuous case is straightforward.

As before, we can use the matrix $e_{xy}$ to define a measure of
assortativity.  We first note that $e_{xy}$ satisfies the sum rules
\begin{equation}
\sum_{xy} e_{xy} = 1,\qquad \sum_y e_{xy} = a_x,\qquad \sum_x e_{xy} = b_y,
\label{sumrulesxy}
\end{equation}
where $a_x$ and $b_y$ are, respectively, the fraction of edges that start
and end at vertices with values $x$ and~$y$.  (On an undirected, unipartite
graph, $a_x=b_x$.)  Then, if there is no assortative mixing
$e_{xy}=a_xb_y$.  If there is assortative mixing one can measure it by
calculating the standard Pearson correlation coefficient thus:
\begin{equation}
r = {\sum_{xy} xy(e_{xy}-a_xb_y)\over\sigma_a\sigma_b},
\label{rscalar}
\end{equation}
where $\sigma_a$ and $\sigma_b$ are the standard deviations of the
distributions $a_x$ and $b_y$.  The value of~$r$ lies in the range $-1\le
r\le1$, with $r=1$ indicating perfect assortativity and $r=-1$ indicating
perfect disassortativity (i.e.,~perfect negative correlation between $x$
and~$y$).  For the age data from Fig.~\ref{age}, for example, we find that
$r=0.574$, indicating strong assortative mixing once more.

One can construct in a straightforward manner a random graph model of a
network with this type of mixing exactly analogous to the model presented
in Section~\ref{modelsdisc}.  It is also possible to generate random
representative networks from the ensemble defined by~$e_{xy}$ using the
algorithm described in Section~\ref{simdisc}.  In this paper however,
rather than working further on the general type of mixing described here,
we will concentrate on one special example of assortative mixing by a
scalar property which is particularly important for many of the networks we
are interested in, namely mixing by vertex degree.

\subsection{Mixing by vertex degree}
\label{bydegree}
In general, scalar assortative mixing of the type described above requires
that the vertices of the network of interest have suitable scalar
properties attached to them, such as age or income in social networks.  In
many cases, however, data are not available for these properties to allow
us to assess whether the network is assortatively mixed.  But there is one
scalar vertex property that is always available for every network, and that
is vertex degree.  So long as we know the network structure we always know
the degree of a vertex, and then we can ask whether vertices of high degree
preferentially associate with other vertices of high degree.  Do the
gregarious people hang out with other gregarious people?  This has been a
topic of considerable discussion in the physics
literature~\cite{PVV01,MS02b,DMS02,VW02,BPV02}.  As we will show, many
real-world networks do show significant assortative (or disassortative)
mixing by vertex degree.

Assortative mixing by degree can be quantified in exactly the same way as
for other scalar properties of vertices, using Eq.~\eref{rscalar}.  Taking
the example of an undirected network and using the notation of
Ref.~\onlinecite{Newman02f}, we define $e_{jk}$ to be the fraction of edges
that connect vertices of degrees $j$ and~$k$.  In fact, we choose $j$
and~$k$ to be the \textit{excess degrees} of the vertices (also called
\textit{remaining degree} in Ref.~\onlinecite{Newman02f}), which are one
less than the degrees of the vertices themselves.  This is because in most
cases we are interested in the number of edges attached to a vertex other
than the particular edge we are looking at at the moment.

If the degree distribution of the graph as a whole is~$p_k$, i.e.,~$p_k$ is
the probability that a randomly chosen vertex will have degree~$k$, then
the excess degree of the vertex at the end of an edge is distributed
according to~\cite{NSW01}
\begin{equation}
q_k = {(k+1) p_{k+1}\over z},
\label{defsqk}
\end{equation}
where $z=\sum_k k p_k$ is the mean degree in the network.  The distribution
$q_k$ is related to $e_{jk}$ via
\begin{equation}
\sum_j e_{jk} = q_k.
\label{degreesum}
\end{equation}
The correct assortativity coefficient for mixing by vertex degree in an
undirected network is
\begin{equation}
r = {\sum_{jk} jk(e_{jk}-q_jq_k)\over\sigma_q^2},
\end{equation}
where $\sigma_q$ is the standard deviation of the distribution~$q_k$.  For
a directed network the equivalent expression is
\begin{equation}
r = {\sum_{jk} jk(e_{jk}-q^{\rm in}_jq^{\rm out}_k)\over
     \sigma_{\rm in}\sigma_{\rm out}},
\label{rdir}
\end{equation}
where $e_{jk}$ is now the probability that a randomly chosen directed edge
leads into a vertex of in-degree~$j$ and out of a vertex of
out-degree~$k$~\footnote{One can also calculate a value for~$r$ by simply
ignoring the directed nature of the edges in a directed network.  This
approach, which we adopted in Ref.~\onlinecite{Newman02f}, will in general
give a different figure from that given by Eq.~\eref{rdir}.  While
Eq.~\eref{rdir} will normally give a more meaningful result for a directed
network, there may be cases in which ignoring direction is the correct
thing to do.  For example, in a food web one might only be interested in
which species have tropic relations with with others, and not in which
direction that relation lies in terms of energy or carbon flow.}.  For the
purposes of calculating~$r$ for an actual network of specified vertices and
edges, we can rewrite this in the form
\begin{equation}
r = {\sum_i j_i k_i - M^{-1} \sum_i j_i \sum_{i'} k_{i'}\over
     \sqrt{\Bigl[\sum_i j_i^2 - M^{-1} \bigl(\sum_i j_i\bigr)^2\Bigr]
           \Bigl[\sum_i k_i^2 - M^{-1} \bigl(\sum_i k_i\bigr)^2\Bigr]}},
\end{equation}
where $j_i$ and $k_i$ are the excess in-degree and out-degree of the
vertices that the $i$th edge leads into and out of respectively, and $M$ is
again the number of edges.  For an undirected network we can use the same
formula---we simply replace each undirected edge by two directed ones
leading in opposite directions.  Alternatively, one can apply directly the
formula given in Ref.~\onlinecite{Newman02f}, Eq.~(4), to the undirected
network.  As before, errors on the measured values of~$r$ can be calculated
using the jackknife method and Eq.~\eref{jackknife}.

\begin{table*}[t]
\hbox to\textwidth{\hspace{25pt}
\begin{tabular}{rl|c|r|c|l|c}
& network                    & type       & size $n$      & assortativity $r$ & error $\sigma_r$ & ref. \\
\cline{2-7}
\multirow{7}{28pt}{$\mbox{social}\left\lbrace\rule{0pt}{37pt}\right.$}
& physics coauthorship       & undirected & $52\,909$     & $\phm0.363$       & $0.002$          & a    \\
& biology coauthorship       & undirected & $1\,520\,251$ & $\phm0.127$       & $0.0004$         & a    \\
& mathematics coauthorship   & undirected & $253\,339$    & $\phm0.120$       & $0.002$          & b    \\
& film actor collaborations  & undirected & $449\,913$    & $\phm0.208$       & $0.0002$         & c    \\
& company directors          & undirected & $7\,673$      & $\phm0.276$       & $0.004$          & d    \\
& student relationships      & undirected & $573$         & $-0.029$          & $0.037$          & e    \\
& email address books        & directed   & $16\,881$     & $\phm0.092$       & $0.004$          & f    \\
\cline{2-7}
\multirow{4}{57.4pt}{$\mbox{technological}\left\lbrace\rule{0pt}{22pt}\right.$}
& power grid                 & undirected & $4\,941$      & $-0.003$          & $0.013$          & g    \\
& Internet                   & undirected & $10\,697$     & $-0.189$          & $0.002$          & h    \\
& World-Wide Web             & directed   & $269\,504$    & $-0.067$          & $0.0002$         & i    \\
& software dependencies      & directed   & $3\,162$      & $-0.016$          & $0.020$          & j    \\
\cline{2-7}
\multirow{5}{43.4pt}{$\mbox{biological}\left\lbrace\rule{0pt}{26pt}\right.$}
& protein interactions       & undirected & $2\,115$      & $-0.156$          & $0.010$          & k    \\
& metabolic network          & undirected & $765$         & $-0.240$          & $0.007$          & l    \\
& neural network             & directed   & $307$         & $-0.226$          & $0.016$          & m    \\
& marine food web            & directed   & $134$         & $-0.263$          & $0.037$          & n    \\
& freshwater food web        & directed   & $92$          & $-0.326$          & $0.031$          & o    \\
\end{tabular}\hfil}
\caption{Size~$n$, degree assortativity coefficient~$r$, and expected error
$\sigma_r$ on the assortativity, for a number of social, technological, and
biological networks, both directed and undirected.  Social networks:
coauthorship networks of (a)~physicists and biologists~\cite{Newman01a} and
(b)~mathematicians~\cite{GI95}, in which authors are connected if they have
coauthored one or more articles in learned journals; (c)~collaborations of
film actors in which actors are connected if they have appeared together in
one or more movies~\cite{WS98,ASBS00}; (d)~directors of Fortune 1000
companies for 1999, in which two directors are connected if they sit on the
board of directors of the same company~\cite{DYB01}; (e)~romantic (not
necessarily sexual) relationships between students at a US high
school~\cite{BMS02}; (f)~network of email address books of computer users
on a large computer system, in which an edge from user~A to user~B
indicates that B appears in A's address book~\cite{NFB02}.  Technological
networks: (g)~network of high voltage transmission lines in the Western
States Power Grid of the United States~\cite{WS98}; (h)~network of direct
peering relationships between autonomous systems on the Internet, April
2001~\cite{Chen02}; (i)~network of hyperlinks between pages in the
World-Wide Web domain \texttt{nd.edu}, \textit{circa} 1999~\cite{AJB99};
(j)~network of dependencies between software packages in the GNU/Linux
operating system, in which an edge from package~A to package~B indicates
that A relies on components of B for its operation.  Biological networks:
(k)~protein--protein interaction network in the yeast
\textit{S. Cerevisiae}~\cite{Jeong01}; (l)~metabolic network of the
bacterium \textit{E. Coli}~\cite{Jeong00}; (m)~neural network of the
nematode worm \textit{C. Elegans}~\cite{WSTB86,WS98}; tropic interactions
between species in the food webs of (n)~Ythan Estuary,
Scotland~\cite{HBR96} and (o)~Little Rock Lake,
Wisconsin~\cite{Martinez91}.}
\label{dctab}
\end{table*}

In Table~\ref{dctab} we show the measured values of~$r$ for degree
correlations in undirected and directed networks of a variety of different
types, along with the expected errors on these values.  The table reveals
an interesting feature: essentially all the social networks examined are
significantly assortative by degree, i.e.,~high degree vertices tend to be
connected to other high degree vertices, while all the technological and
biological networks are disassortative.  Three of the values for~$r$, for
the network of student relationships, the power grid, and the graph of
software dependencies, are null results, meaning that they are not
statistically different from zero.  All the others however fit the pattern
clearly, with positive values of~$r$ for the social networks and negative
values for all the others.

What is the explanation of this phenomenon?  In all probability, there are
a number of different mechanisms at work.  Some possibilities are the
following:
\begin{enumerate}
\item In the social networks it is entirely possible, and is often assumed
in the sociological literature, that similar people do attract one another,
and therefore that there could be a real preference among gregarious people
for association with other gregarious people, and similarly for hermits.
\item On the other hand, the networks of collaborations between academics,
actors, and businesspeople considered here are affiliation networks,
i.e.,~networks in which people are connected together by membership of
common groups (authors of a paper, actors in a film, etc.).  Since all
members of a group are thus connected to all other members, the positive
correlations between degrees may at least in part reflect the fact that the
members of a large (small) group are connected to the other members of the
same large (small) group.
\item In the Internet and the World-Wide Web there may be organizational
reasons for degree anti-correlation between vertices.  The high-degree
vertices in these networks are often connectivity providers (Internet) or
directories (Web), which by definition tend to be connected to the ``little
people''---the individual service subscribers in the case of the Internet
or the individual web-pages on the Web.
\item Maslov and Sneppen~\cite{MS02b} have shown that disassortativity can
be produced as a finite-size effect by the constraint that no two vertices
in a network are connected by more than one edge.  This constraint causes
high-degree vertices to ``repel'' one another, producing negative values
of~$r$.  This explanation could account for at least a part of the
disassortative mixing we see in the Internet, the protein and metabolic
networks, and the food webs, although it cannot be applied directly to the
Web and neural networks, for which vertex pairs can and often do have more
than one connection.
\end{enumerate}
It appears therefore that some of the degree correlations we see in our
networks could have real social or organizational origins, while others may
be artifacts of the types of networks we are looking at and the constraints
that are placed on their structure.

\subsection{Models of assortative mixing by degree}
\label{secmc}
In Ref.~\onlinecite{Newman02f} we studied the ensemble of graphs that have
a specified value of the matrix~$e_{jk}$ and solved exactly for its average
properties using generating function methods similar to those of
Section~\ref{modelsdisc}.  We showed that the phase transition at which a
giant component first appears in such networks occurs at a point given by
$\det(\vI-\vm)=0$, where $\vm$ is the matrix with elements $m_{jk} =
ke_{jk}/q_j$.  One can also calculate exactly the size of the giant
component, and the distribution of sizes of the small components below the
phase transition.  While these developments are mathematically elegant,
however, their usefulness is limited by the fact that the generating
functions involved are rarely calculable in closed form for arbitrary
specified~$e_{jk}$, and the determinant of the matrix~$\vI-\vm$ almost
never is.  In this paper, therefore, we take an alternative approach,
making use of computer simulation.

We would like to generate on a computer a random network having, for
instance, a particular value of the matrix~$e_{jk}$.  (This also fixes the
degree distribution, via Eq.~\eref{degreesum}.)  In
Ref.~\onlinecite{Newman02f} we discussed one possible way of doing this
using an algorithm similar that of Section~\ref{simdisc}.  One would draw
edges from the desired distribution~$e_{jk}$ and then join the degree~$k$
ends randomly in groups of~$k$ to create the network.  (This algorithm has
also been discussed recently by Dorogovtsev~\etal~\cite{DMS02}.)  As we
pointed out, however, this algorithm is flawed because in order to create a
network without any dangling edges the number of degree~$k$ ends must be a
multiple of~$k$ for all~$k$.  It is very unlikely that these constraints
will be satisfied by chance, and there does not appear to be any simple way
of arranging for them to be satisfied without introducing bias into the
ensemble of graphs.  Instead, therefore, we use a Monte Carlo sampling
scheme which is essentially equivalent to the Metropolis--Hastings method
widely used in the mathematical and social sciences for generating model
networks~\cite{Strauss86,Snijders02}.  The algorithm is as follows.
\begin{enumerate}
\item Given the desired edge distribution $e_{jk}$, we first calculate the
corresponding distribution of excess degrees~$q_k$ from
Eq.~\eref{degreesum}, and then invert Eq.~\eref{defsqk} to find the degree
distribution:
\begin{equation}
p_k = {q_{k-1}/k\over\sum_j q_{j-1}/j}.
\end{equation}
Note that this equation cannot tell us how many vertices there are of
degree zero in the network.  This information is not contained in the edge
distribution~$e_{jk}$ since no edges connect to degree-zero vertices, and
so must be specified separately.  On the other hand, most of the properties
of networks with which we will be concerned here don't depend on the number
of degree-zero vertices, so we can safely set $p_0=0$ for the purposes of
this paper.
\item We draw a degree sequence, a specific set $k_i$ of degrees of the
vertices $i=1\ldots N$, from the distribution~$p_k$, and connect vertices
together randomly in pairs to generate a random graph, as described, for
instance, by Molloy and Reed~\cite{MR95}.
\item We choose two edges at random from the graph.  Let us denote these by
the vertex pairs $(v_1,w_1)$ and $(v_2,w_2)$ that they connect.
\item We measure the excess degrees $j_1,k_1,j_2,k_2$ of the vertices
$v_1,w_1,v_2,w_2$ and then we remove the two edges and replace them with
two new ones $(v_1,v_2)$ and $(w_1,w_2)$ with probability
\begin{equation}
P = \Biggl\lbrace\begin{array}{ll}
      \phantom{\bigg|}{\displaystyle e_{j_1j_2} e_{k_1k_2}\over
                      \displaystyle e_{j_1k_1} e_{j_2k_2}} &
      \qquad\mbox{if $e_{j_1j_2} e_{k_1k_2}<e_{j_1k_1} e_{j_2k_2}$} \\
      \phantom{\big|}1 & \qquad\mbox{otherwise.}
    \end{array}
\label{acceptance}
\end{equation}
\item Repeat from step~3.
\end{enumerate}
Clearly this swap procedure preserves the degree sequence.  It is also
ergodic over the set of graphs with that degree sequence, i.e.,~it can
reach any graph within that set in a finite number of moves.  To see this,
consider any configuration of the graph other than the desired target
configuration and choose any vertex pair that is not joined by an edge in
that configuration but \emph{is} joined by an edge in the target.  These
vertices must necessarily each be attached to at least one other edge that
does not exist in the target configuration.  Take these edges and perform
the swap procedure on them.  This always increases the number of edges that
the configuration has in common with the target.  And since it is always
possible to do this, it immediately follows that any target configuration
can be reached in at most $M$ such moves, where $M$ is the number of edges
in the network.  (Actually, $M-1$ will suffice, since the last edge will
always automatically be in the correct position by a process of
elimination.)

Our algorithm also satisfies detailed balance.  We would like to sample
graph configurations~$\mu$ with probabilities~\footnote{Strictly these
probabilities are only correct in a ``canonical ensemble'' of graphs in
which the degree distribution is fixed rather than the degree sequence.
This ensemble and the fixed-degree-sequence one studied here, however,
become equivalent in the limit of large graph size; the error introduced
here by substituting one for the other is of the order of~$N^{-1}$ and is
small compared with other sources of error in our simulations.}
\begin{equation}
p_\mu = \prod_{i=1}^M e_{j_ik_i},
\label{defspmu}
\end{equation}
where $j_i,k_i$ are the excess degrees of the vertices at the ends of the
$i$th edge.  It is trivial to show that with the choice of transition
probabilities given in Eq.~\eref{acceptance}, we satisfy detailed balance
in the form
\begin{equation}
p_\mu P(\mu\to\nu) = p_\nu P(\nu\to\mu),
\end{equation}
for all pairs $\mu,\nu$ of states.

Since our algorithm satisfies both ergodicity and detailed balance, it
immediately follows~\cite{NB99} that in the limit of long time it samples
graph configurations correctly from the distribution~\eref{defspmu}.  It
also turns out to be a reasonably efficient algorithm in practice.  In the
simulations reported here, the mean fraction of proposed Monte Carlo moves
that was accepted never fell below 50\% for any parameter values.

To use this algorithm we also need to choose a value for the
matrix~$e_{jk}$.  We have a lot of freedom about how we do this.  Suppose
for example that we wish to simulate an undirected network, so that
$e_{jk}$ is symmetric.  A rank-$n$ symmetric matrix has $\half n(n+1)$
degrees of freedom, $n$~of which are fixed in this case by the requirement,
Eq.~\eref{degreesum}, that the rows and columns sum to~$q_k$, leaving
$\half n(n-1)$ that can be freely chosen.  (One could think for example of
choosing the $\half n(n-1)$ off-diagonal elements of $e_{jk}$ and then
satisfying~\eref{degreesum} by choosing the $n$ diagonal elements to give
the row and column sums the desired values.)

A simple example of a disassortative choice for $e_{jk}$ is the set of
matrices taking the form
\begin{equation}
e_{jk}^{(d)} = q_j x_k + x_j q_k - x_j x_k,
\label{simple}
\end{equation}
where $x_k$ is any distribution normalized such that $\sum_k x_k=1$.  It is
easy to see that this choice satisfies the constraints on $e_{jk}$, and the
resulting value of~$r$ is
\begin{equation}
r_d = -{(\mu_q-\mu_x)^2\over\sigma_q^2},
\end{equation}
which is always negative (or zero).  Here $\mu_q$ and $\mu_x$ are the means
of the distributions $q_k$ and $x_k$ respectively.

Being a probability, $e_{jk}$~is also constrained to lie in the range $0\le
e_{jk}\le1$ for all~$j,k$.  To ensure that Eq.~\eref{simple} never becomes
negative we should choose $x_k$ to decay faster than~$q_k$.

\begin{figure}
\resizebox{7cm}{!}{\includegraphics{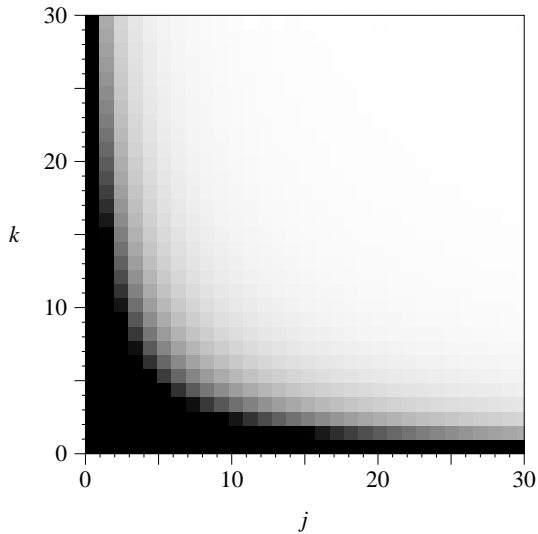}}
\caption{Density plot of the matrix $e_{jk}^{(d)}$, Eq.~\eref{simple}, for
$\tau=2.5$, $\kappa=100$, $\kappa'=10$.  Darker squares represent higher
values of~$e_{jk}$.}
\label{density}
\end{figure}

Suppose for example that we are interested, as many people seem to be these
days, in networks that have power-law degree distributions, $p_k\sim
k^{-\tau}$~\cite{Strogatz01,AB02,DM02,Price65,Redner98,AJB99,FFF99,ASBS00,ACL00,Liljeros01,EMB02}.
True power laws unfortunately are troublesome to deal with; the crucial
distribution $q_k$ has a divergent mean unless~$\tau>3$, which it seldom is
for real-world networks (see, for instance, Ref.~\onlinecite{AB02},
Table~II).  Instead, therefore, following Ref.~\onlinecite{NSW01}, we here
examine the exponentially truncated power-law distribution
\begin{equation}
p_k = {k^{-\tau} \e^{-k/\kappa}\over\Li_\tau(\e^{-1/\kappa})} \qquad
\mbox{for $k\ge1$,}
\label{truncated}
\end{equation}
where the function $\Li_n(x)$, which acts here as a normalizing constant,
is the $n$th polylogarithm of~$x$.  This gives a similar distribution
$q_k\sim(k+1)^{-\tau+1} \e^{-(k+1)/\kappa}$ for the excess degree, and we
choose $x_k$ to have the same functional form, but with a different cutoff
parameter $\kappa'$, where $\kappa'<\kappa$, to ensure that~$e_{jk}>0$.  In
Fig.~\ref{density} we show a plot of the resulting $e_{jk}$ for $\tau=2.5$,
$\kappa=100$, $\kappa'=10$.  The disassortative nature of this choice for
$e_{jk}$ is evident from the concentration of probability along the edges
of the matrix in the figure.

Introducing a cutoff in the degree distribution also provides us with a
parameter, namely~$\kappa$, that can, as we will shortly see, be
conveniently manipulated to produce a phase transition at which a giant
component appears in the network.

We can also make an assortative matrix $e_{jk}$ by writing
\begin{equation}
e_{jk}^{(a)} = 2q_jq_k - e_{jk}^{(d)},
\end{equation}
which gives a value for the assortativity coefficient of $r_a=-r_d$, while
still having the same degree distribution.  More generally, we would like
to be able to vary $r$ freely, keeping the degree distribution fixed.  We
can do this by writing $e_{jk}$ in the form
\begin{equation}
e_{jk} = q_j q_k + r\sigma_q^2 m_{jk},
\label{simejk}
\end{equation}
where the symmetric matrix $m_{jk}$ has all row and column sums zero and is
normalized such that
\begin{equation}
\sum_{jk} jkm_{jk} = 1.
\end{equation}
For any choice of $m_{jk}$ satisfying these constraints, Eq.~\eref{simejk}
gives us a one-parameter family of networks parametrized by the
assortativity coefficient~$r$.  We can for example choose
\begin{equation}
m_{jk} \sim q_jq_k - e_{jk}^{(d)} = (q_j - x_j)(q_k - x_k),
\label{simmjk}
\end{equation}
for any correctly normalized~$x_k$.  Then Eq.~\eref{simejk} allows us to
interpolate smoothly between $e_{jk}^{(d)}$ and $e_{jk}^{(a)}$ (and beyond)
by simply varying the value of~$r$.  Note that whenever $r=0$, we get a
simple random graph without degree correlations, of the type discussed by
Molloy and Reed~\cite{MR95} and others~\footnote{Not all graphs with $r=0$
are without degree correlations.  A measurement of $r=0$ implies only that
the mean degree correlation is zero when averaged over all degrees.  The
grown graph model of Barab\'asi and Albert~\cite{BA99b} provides an example
of a network that possesses degree correlations although it has
$r=0$~\cite{Newman02f}.}.

\subsection{Simulation results}
\label{simresults}
For $e_{jk}$ chosen according to Eqs.~\eref{simejk} and~\eref{simmjk}, with
$q_k$ and $x_k$ taking the same truncated power-law form as in
Fig.~\ref{density}, we have performed simulations for a variety of values
of the parameters $\tau$, $\kappa$, and~$\kappa'$.

\begin{figure}
\resizebox{\figurewidth}{!}{\includegraphics{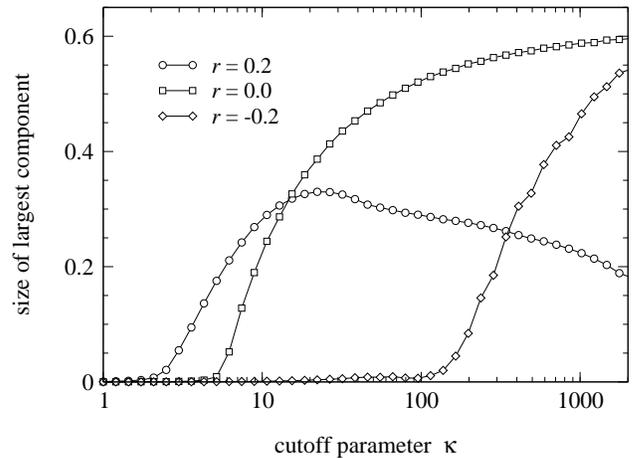}}
\caption{Monte Carlo simulation results for the size of the giant component
in graphs with the degree distribution~\eref{truncated}, as a function of
the cutoff parameter~$\kappa$, with $\tau=2.5$ and $\kappa'=2$.  The
individual curves are for different values of the assortativity, as marked.
Each data point is an average over 100 graphs of $100\,000$ vertices.}
\label{gc}
\end{figure}

In Fig.~\ref{gc} we show the size of the largest component in the graph as
a function of~$\kappa$, for three different values of the assortativity
coefficient~$r$.  As the cutoff parameter $\kappa$ increases, the mean
degree of the graph increases also, so that the graph becomes more dense,
ultimately passing the critical point at which a giant component develops.
The figure reveals two findings of particular note.
\begin{enumerate}
\item The position of the phase transition at which the giant component
appears moves to higher values of~$\kappa$ as the value of $r$ decreases.
That is, the more assortative a network is, the lower the edge density at
which a giant component first appears.  This is intuitively reasonable.  In
assortatively mixed networks, the high-degree vertices tend to associate
preferentially with one another, sticking together and forming what in the
epidemiological literature is called a ``core group.''  Within this core
group the edge density is higher than it is in the graph as a whole, since
the vertices in the group have higher-than-average degree.  Thus one would
expect to see a giant-component forming in the core group before it would
form in a graph of the same average density but with no assortative mixing.
Conversely, in graphs that are disassortatively mixed, the phase transition
happens at a higher density than in neutrally mixed graphs.
\item The size of the giant component in the limit of large~$\kappa$ is
smaller for the assortatively mixed graph than for the neutral and
disassortative ones.  While this might seem at first to be at odds with the
result that assortative graphs show a phase transition at lower density, it
is really a reflection of the same underlying mechanism.  Although the
presence of a core group in an assortative graph allows it to percolate at
a lower average density than other graphs, it also means that the density
in other parts of the graph, outside the core group, is lower, and hence
that the giant component is unlikely to extend into those regions.  Thus
the giant component is confined to the core of the network, and cannot grow
as large as in a neutral or disassortative network.
\end{enumerate}

A question of considerable interest in the study of networked systems is
that of network resilience to the deletion of vertices.  Suppose vertices
are removed one by one from a network.  How many must be removed before the
giant component of the network is destroyed and network communication
between distant vertices can no longer take place?  Many networks,
particularly those with highly skewed degree distributions, are found to be
resilient to the random deletion of vertices but susceptible to the
targeted deletion specifically of those vertices that have the highest
degrees~\cite{AJB00,Broder00,CNSW00,CEBH01}.  As we now show, these general
results are modified by the presence of assortative mixing in the network.

\begin{figure}
\resizebox{\figurewidth}{!}{\includegraphics{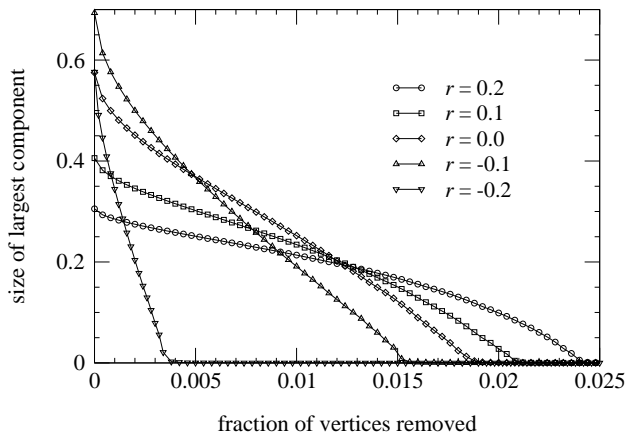}}
\caption{The size of the largest component in a network as a fraction of
system size as the highest-degree vertices are removed one by one.  Each
curve is for a single network of $10^7$ vertices generated using the Monte
Carlo method described in the text, with the degree distribution and the
matrix~$e_{jk}$ chosen according to Eqs.~\eref{simple}
and~\eref{truncated}, with $\tau=2.5$, $\kappa=500$, $\kappa'=5$, and
$r$-values as marked.  Clearly the most assortatively mixed network
($r=0.2$) is considerably more robust against the removal of vertices than
the most disassortative ($r=-0.2$).}
\label{resil}
\end{figure}

In Fig.~\ref{resil}, we show the size of the largest component for five
networks with different values of~$r$ as vertices are removed in decreasing
order of their degree---i.e.,~highest degree vertices first~\footnote{The
degree is not recalculated after each removal.  Removal is in the order of
vertices' starting degree in the network before any deletion has taken
place.}.  As the figure shows, although each of the networks has the same
degree distribution, there is dramatic variation in the resilience of the
networks with their assortativity.  For the most assortative network, with
$r=0.2$, it requires the removal of about ten times as many high-degree
vertices to destroy the giant component as for the most disassortative one,
with $r=-0.2$, even though the disassortative network starts out with a
giant component about twice as big.

This finding fits naturally with our picture of an assortative network as
dominated by a core group of interconnected high-degree vertices.  Such a
core group provides robustness to the network by concentrating all the
obvious target vertices together in one portion of the graph.  Removing
these high-degree vertices is still one of the most effective ways to
destroy network connectivity, but it is less effective because by removing
them all from the same portion of the graph we fail to attack other
portions.  And if those other portions are themselves percolating, then a
giant component will persist even as the highest degree vertices vanish.

Conversely, the disassortatively mixed network is particularly susceptible
to removal of high-degree vertices because those vertices are strewn far
apart across the network, so that attacking them attacks all parts of the
network at once.

One can also ask about the resilience of networks under random failure of
their vertices (rather than targeted attack)~\cite{AJB00,CEBH00,CNSW00}.
Although we do not treat this case in detail here, it is reasonable to
suppose that it is similar to the case of targeted attack.  If assortative
mixing makes networks more resilient against removal of their
highest-degree vertices, then presumably they will also be resilient
against removal of random ones; random vertex failure will do most damage
when it happens to hit high-degree vertices, but, as we have seen, this
vulnerability is diminished by the concentration of the high-degree
vertices in the core group.  Some qualitative behaviors of the system may
be unaffected by assortativity, however.  For example, it is known that the
fraction of vertices that must randomly fail to destroy the giant component
in a network with a power-law degree distribution and uncorrelated degrees
tends to unity as graph size becomes large, provided the exponent of the
power law satisfies $\tau<3$~\cite{CEBH00}.  Vazquez and
Moreno~\cite{VM03} have recently shown that this result is not affected by
the presence of assortative mixing by degree in the network, although
disassortative mixing can make a difference.

\subsection{Discussion}
The results found here could have applications in a variety of areas.
Consider for example, the spread of diseases on networks, which has been
the subject of much attention in the recent networks
literature~\cite{BMS97,Andersson99,MN00a,PV01a,PV01b,ML01,Sander02,Newman02c}.
The largest component of the contact network over which a disease spreads
represents the largest possible disease outbreak on that network, and a
network with no giant component cannot show an epidemic (system-wide)
outbreak.  Thus our finding that a giant component forms more easily in a
network that is assortatively mixed by degree suggests that in such
networks epidemic outbreaks would become possible at a lower edge density
than in the corresponding disassortative network.  In the language of
epidemiology, the core group of an assortatively mixed network forms a
``reservoir,'' which can sustain an outbreak of the disease even when the
density of the network as a whole is too low to do so.  On the other hand,
the smaller asymptotic size of the giant component in an assortatively
mixed network seems to imply that, when they occur, epidemics in such
networks would be restricted to a smaller segment of the population than in
a similar disassortative network---the outbreak is confined mostly to the
core group and does not spread to the population as a whole.  Thus from the
epidemiological point of view there are both good and bad sides to the
phenomenon of assortativity.

One could test these predictions explicitly by studying epidemic models
such as SIR or SIRS models~\cite{Bailey75,Hethcote00} on assortatively
mixed model networks of the type introduced here.  Some studies of this
kind have already been carried out---see for example
Refs.~\onlinecite{BPV02} and~\onlinecite{MV03}---although the particular
conjectures put forward here have not been conclusively verified.

Our findings on network resilience also have some practical implications.
In the context of epidemiology, for instance, removal of vertices from the
network might correspond to immunization of individuals to prevent the
spread of disease.  Assuming that the goal of a vaccination program is to
destroy network connectivity so that the disease in question cannot spread,
our findings suggest that even targeted vaccination strategies would be
less effective in assortative networks than in disassortative or neutral
ones because of the resilience of the network to this type of attack.  In
other contexts, however, resilience is a good thing.  For example we would
like to make networks such as the Internet and other communication or
distribution networks resilient against attacks on their vertices.  In this
context assortative mixing would be beneficial.

Unfortunately, when we look at Table~\ref{dctab}, we find a discouraging
picture.  As we pointed out in Section~\ref{bydegree}, almost all the
social networks we have looked at are significantly assortative, meaning
that they would be robust to vertex removal.  But these are the very
networks by which disease spreads, the ones that we would like to be able
to attack using vaccination strategies.  Even the email network, which is
relevant to the spread of computer viruses~\cite{NFB02}, is assortative and
hence resilient.  On the other hand, the technological networks like the
Internet, which we would like to be able to protect, are disassortative,
and hence particularly vulnerable to targeted attack.

\section{Conclusions}
\label{concs}
In this paper we have studied the phenomenon of assortative mixing in
networks, which is the tendency for vertices in networks to connect
preferentially to other vertices that are like them in some way.  This
preference may take a number of forms.  Mixing may follow discrete or
enumerative characteristics.  In the social networks that have been the
main focus of this paper, connections between people may be assortative by
language, for example, or by race---people may prefer to associate with
others who speak the same language as they do or are of the same race.
Mixing can also be dictated by scalar characteristics such as age or
income.  A special case of mixing by a scalar characteristic is mixing
according to vertex degree, which has been shown previously to be present
in a variety of networks, including non-social ones such as the Internet
and protein interaction networks.  Mixing can also be disassortative,
meaning that vertices in the network preferentially form connections to
others that are unlike them.

We have proposed some simple measures for these types of mixing, which we
call assortativity coefficients.  These measures are positive or negative
for assortative or disassortative mixing respectively, and zero for
neutrally mixed networks.  Applying our measures to a broad selection of
network data drawn from various real-world situations we have shown that
the phenomenon of assortative mixing is indeed widespread, with only a few
of the networks studied showing no statistically significant biases in
their mixing patterns.  In the case of mixing by vertex degree, a
remarkable pattern emerges.  Almost all the social networks studied show
positive assortativity coefficients while all other types of networks,
including technological and biological networks, show negative
coefficients, i.e.,~disassortative mixing.  Only three networks that showed
no significant trend either way failed to follow this rule.  We have
offered some conjectures about the origin of this striking regularity, but
we believe it unlikely that any single mechanism can explain the mixing
patterns of all of these disparate networks.

We have also proposed a number of models of assortatively mixed networks,
for mixing both by discrete and by scalar characteristics.  For each of the
mixing types considered it is possible to create random graph models for
which one can calculate exactly by generating function methods certain
average properties of network ensembles.  We have also described Monte
Carlo methods for generating random graphs drawn from each of the classes
discussed with specified values of the mixing parameters.

For the case of mixing by vertex degree we have performed extensive
simulations.  Two results of particular interest emerge from these studies.
First, we find that networks that are assortatively mixed by degree
percolate more easily that their disassortative counterparts.  That is, a
giant component of connected vertices forms in the network at lower edge
density than in another network with the same degree distribution but zero
or negative assortativity.  This result may imply, for instance, that
assortatively mixed social networks would support epidemic disease
outbreaks more easily than disassortatively mixed ones, which would be a
disheartening conclusion, given our finding that most social networks
appear to be assortative.

Second, we find that assortatively mixed networks are more robust to the
deletion of their vertices than disassortatively mixed or neutral networks.
We have studied in particular the case of the targeted deletion of the
highest-degree vertices, which has been suggested as a possible vaccination
strategy for breaking up networks of disease-causing contacts, but it is
reasonable to suppose that the same result will extend also to the random
deletion of vertices.  This result too leads to a rather gloomy conclusion:
targeted vaccination strategies may be less effective than we would hope in
preventing disease because of the assortative and hence resilient nature of
social networks, while on the other hand networks that we would hope to
protect against vertex removal, communication networks like the Internet,
for instance, will be particularly susceptible because of their
disassortative nature.

\begin{acknowledgments}
  The author thanks Michelle Girvan, Richard Rothenberg, and Matthew
  Salganik for useful conversations and L\'aszl\'o Barab\'asi, Jerry Davis,
  Jennifer Dunne, Jerry Grossman, Hawoong Jeong, Neo Martinez, Duncan
  Watts, and the Inter-University Consortium for Political and Social
  Research for providing data used in the calculations.  This work was
  supported in part by the National Science Foundation under grants
  DMS--0109086 and DMS--0234188.
\end{acknowledgments}

\appendix
\section{The assortativity measure of Gupta~et~al.}
\label{appa}
Gupta~\etal~\cite{GAM89} have defined a measure of assortative mixing by
discrete types different from the one that we have introduced in
Section~\ref{discmeas}.  In our notation their measure is
\begin{equation}
Q = {\sum_i a_i^{-1} (e_{ii} - a_i b_i)\over n-1}
  = {\sum_i e_{ii}/a_i - 1\over n - 1},
\end{equation}
where as before $n$ is the number of vertex types, and we have made use of
$\sum_i b_i = 1$.  Like our measure, this measure is 0 for a neutrally
mixed network and 1 for a perfectly assortative network.  In general,
however, the values of the two measures are quite different.  Here we give
a simple example to illustrate the difference between the two.

Consider a network with three types of vertices.  There are 100 vertices of
type~1, 100 of type~2, and 2 of type~3.  The vertices of types~1 and~2 mix
indiscriminately with one other---connections from 1 to 2 are as likely as
from 1 to 1, and so forth.  The vertices of type~3 however associate only
among themselves and not with types 1 and 2 at all.  This is reflected in
the matrix $\ve$ for the 202 vertices, which for a network with mean
degree~2 would look like this:
\begin{equation}
\ve = {1\over202}
      \begin{pmatrix}
        50 & 50 & 0 \\
        50 & 50 & 0 \\
         0 &  0 & 2 \\
      \end{pmatrix}.
\end{equation}
Clearly most of this network---99\% of it, in fact---is mixing randomly,
and hence we would expect the assortativity coefficient to be close to
zero.  The value of $r$ for the matrix above reflects this; we find
$r=0.029$.  The measure of Gupta~\etal~\cite{GAM89}, however, takes a value
$Q=0.50$.  This appears to indicate that the network has very strong
assortative mixing, when in fact it does not.  The reason for this is that
the measure of Gupta~\etal, rather than giving each vertex in the network
equal weight, weights each \emph{type} of vertex equally, so that vertices
that belong to large groups get less weight in the calculation than those
in small groups.  In the present case, where one group is very small, the
vertices in that group are weighted very heavily, and since those vertices
mix perfectly assortatively, the value of $Q$ is, as a result, high.  If we
remove these vertices from the network, the value of Gupta~\etal's $Q$
coefficient jumps to zero.  Thus the two vertices in the third group have a
disproportionately large effect on the value of~$Q$.

The solution to this problem is to give each vertex equal weight in the
calculation, which is precisely what our measure $r$ does.

\end{document}